\documentclass[aps,prb,showpacs,twocolumn]{revtex4}
\usepackage{graphicx}
\usepackage{bm}
\usepackage{epsf}
\tighten
\newcommand{\beq}{\begin{eqnarray}}
\newcommand{\eeq}{\end{eqnarray}}
\renewcommand{\vec}[1]{{\mathbf{#1}}}
\begin{document}
%\draft
%\twocolumn
%\input epsf.sty
%\input psfig.sty
%\voffset=.75in
\title
{Spin Structure Factor of the Frustrated Quantum 
    Magnet ${\bf {\rm Cs_2 Cu Cl_4 } }$ }
\author{Denis Dalidovich, Rastko Sknepnek, A. John Berlinsky,\\
Junhua Zhang, and Catherine Kallin }
\vspace{.05in}
%
%\begin{instit}
\address
{Department of Physics and Astronomy, McMaster University,\\
Hamilton, Ontario, Canada L8S 4M1 }
\date{\today}

%\end{instit}
%
%\address{\mbox{ }}
%\address{\parbox{14.5cm}{\rm \mbox{ }\mbox{ }
\begin{abstract}
The ground state properties 
and neutron structure factor for
the two-dimensional antiferromagnet on the triangular lattice,
with uni-directional anisotropy in the nearest-neighbor exchange
couplings and a weak Dzyaloshinskii-Moriya (DM) interaction, are studied.  
This Hamiltonian has been used to 
interpret neutron scattering measurements on the
spin $1/2$ spiral spin-density-wave system,
Cs$_2$CuCl$_4$, [R. Coldea, et al.,
Phys. Rev. B {\bf 68}, 134424 (2003)].  
Calculations are performed using a $1/S$ expansion, taking 
into account interactions
between spin-waves. The ground state energy, the shift of the ordering 
wave-vector, $\vec Q$, and the local magnetization are all calculated to 
order $1/S^2$. 
The neutron structure factor, obtained using anharmonic spin-wave
Green's functions to order $1/S$, is shown to be in reasonable agreement with
published neutron data, provided that slightly different parameters
are used for the exchange and DM interactions than those inferred 
from measurements in high magnetic field.
\end{abstract}
%\address{\mbox{ }}
%\address{\mbox{ }}

%\begin{multicols}{2}
%\twocolumn
%\columnseprule 0pt \narrowtext
\maketitle

\section{Introduction}

There is enormous interest in the condensed matter community in the
possibility of observing fractionalized quasiparticles in two or three
dimensional electronic systems.\cite{senthil1,pwa,kivelson}  Fractionalization is the rule, rather
than the exception, in one dimensional metals where electronic spin and
charge propagate independently.\cite{luttinger} 
Fractionally charged quasiparticles have
also been found in the two dimensional electron gas in high
magnetic fields where the fractional quantum Hall 
effect is observed.\cite{FQHE} 
On the theoretical side, fractionalization has recently 
been argued to exist in frustrated two-dimensional quantum model
systems,\cite{spinliq,spinliq1} although no corresponding experimental 
systems have yet been confirmed. A set of long-wavelength theories have
been developed that describe the properties of putative fractionalized 
magnets.\cite{balents1,balents2,balents3}
Subsequently, the theoretical modeling of possible fractionalized
phases was extended to a wide variety of strongly correlated
electron systems.\cite{demler,senthil,spinliq2}
The discovery of fractionalization in real two or three 
dimensional systems in zero magnetic field would provide a
striking example of the emergence of new kinds of particles from the
collective behavior of strongly interacting electrons.
\cite{senthil1}

As a result of this intense interest, the recent claim of the observation
of spinons, neutral, spin 1/2 quasiparticles, in the insulating
antiferromagnet ${\rm Cs_2CuCl_4}$ has attracted 
considerable attention.\cite{coldea,coldea1}
${\rm Cs_2CuCl_4}$ consists of nearly uncoupled layers of 
coupled chains.
The chains are staggered, so that each spin interacts equally with two
spins on each of two neighboring chains, with an antiferromagnetic
coupling, $J^\prime$, about 1/3 the strength of the intrachain
antiferromagnetic nearest neighbor coupling, $J$. The layer thus resembles
a triangular lattice, with strong interactions along one
direction and weaker interactions in the transverse directions. These
intralayer couplings, $J$ and $J^{\prime}$ assisted by a much weaker interlayer
coupling $J^{\prime\prime}$, stabilize an incommensurate spin density wave
(SDW) ground state with a wave vector oriented along the chain direction.
A weak Dzyaloshinskii-Moriya (DM) interaction $D$ is believed to orient the
spins in the 2D layer.

Coldea and coworkers\cite{coldea,coldea1} used neutron scattering to 
study spin-wave excitations from this ordered ground state.  
While Coldea et al. did observe excitations similar 
to what is expected for spin-waves, they also saw broad features which 
they interpreted as excitations of pairs of spinons. Their idea is
that the higher energy spin excitations resemble excitations from a 
nearby-lying spin-liquid state that supports fractionalized
excitations. The low-dimensional nature of the system
together with the low spin are argued to place this system close to a
quantum critical point separating the ordered SDW state from 
a 2D spin liquid state. This conjecture has stimulated a number of theoretical 
studies of possible 2D spin-liquid states that could account for the observed
behavior.\cite{series,series2,largen,slaveboson,slaveboson1,merino,alicea}

In this paper, we pursue a different interpretation of the neutron data,
namely that low spin and quasi-one-dimensionality, along with non-collinear
SDW order, all give rise to substantial anharmonic interactions that
couple one- and two-spinwave excitations from the ordered state.  The
coupling to two spinwave states also yields broad spectra as seen in the
neutron data. A similar conclusion has been reached by Veillette et
al.,\cite{veillette} who also noted that the low momentum resolution 
of the neutron data is another significant factor in the observed 
breadth of the spectra. We find that anharmonic effects lead to 
significant broadening of the neutron spectra but are sensitive to 
anisotropies, such as the DM interaction, which suppress quantum 
fluctuations.  We show in particular that, if such anisotropies 
are weak, then quantum fluctuations, described in terms of a
$1/S$ expansion about the mean field SDW state, reduce the local moment
and lead to substantial
renormalizations and broadening of the excitation spectra.

The rest of this paper is organized as follows.  In Sec. II we describe
the Hamiltonian and the SDW state of ${\rm Cs_2CuCl_4}$.
In Sec. III we show how anharmonic effects modify the ground state energy
and the wave vector $\vec Q$ of the SDW to order $1/S^2$, and
we also review the calculation of the anharmonic one-spin-wave Green's
functions to order $1/S$.  In Section IV we calculate the sublattice 
magnetization $M$ to order $1/S^2$ as a function of the ratio 
$J^\prime /J$ and strength of DM interaction $D/J$, showing that 
the renormalized value of $M$ depends sensitively on the
ratio $D/J$ for small $D/J$. In Sec. V we examine expressions for 
the anharmonic spin-wave energies and damping to order $1/S$.  
For $D=0$ we study how the Goldstone mode at $Q$ is preserved and
show how the preservation of this zero energy mode can be used to define
a set of renormalized coupling constants. Section VI reviews the calculation 
of the neutron structure factor and presents a detailed comparison 
to the inelastic data of Coldea et al. for specific wave vectors 
where two-magnon scattering is important. In Sec. VII we discuss the 
implications of our theoretical calculations in interpreting
existing data on ${\rm Cs_2CuCl_4}$. We also make suggestions for possible 
new experiments.

\section{Spin Density Wave State for ${\rm {\bf Cs_2CuCl_4}}$}

\begin{figure}
\includegraphics[height=3.5cm]{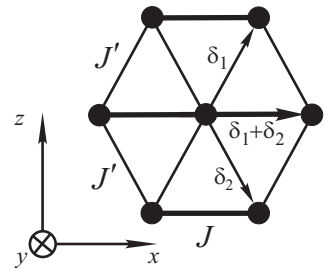}
\caption{Exchange couplings between the different sites of the
triangular lattice within a single layer.}
\label{fig_model}
\end{figure}

The spin density wave state of ${\rm Cs_2CuCl_4}$ can be
described by the simple, model Hamiltonian for a set of decoupled layers:
\begin{eqnarray}
\label{hamil}
H &=& \sum_{\vec R}\big[  J\vec S_{\vec R} \cdot
\vec S_{\vec R +\bm{\delta}_1 +\bm{\delta}_2 } 
+ J^{\prime} \vec S_{\vec R} \cdot (\vec S_{\vec R +\bm{\delta}_1}
+\vec S_{\vec R +\bm{\delta}_2} ) \nonumber \\
 &-& (-1)^n \vec D \cdot \vec S_{\vec R} \times 
(\vec S_{\vec R +\bm{\delta}_1}+\vec S_{\vec R +\bm{\delta}_2}) \big],
\end{eqnarray}
where $J$ is the nearest neighbor coupling constant between $S=1/2$
${\rm Cu}^{2+}$ spins along chains in one direction, which we take to be the
$x$-direction in a triangular lattice, while $J^\prime$, is the coupling
constant along the other two principal directions in each layer
as illustrated in Fig.~\ref{fig_model}. Both $J$ and $J^\prime$ are 
antiferromagnetic, and $J > J^\prime$. The last term in Eq. 
(\ref{hamil}) with $\vec D =(0,D,0)$ describes the
Dzyaloshinskii-Moriya interaction that alternates in sign
between even and odd layers labeled by index $n$.
 
Values of the coupling constants, determined from measurements 
in high magnetic fields,\cite{coldea2} are:
\beq\label{J}
J=0.374\ {\rm meV},\ \ 
\label{Jprime}
J^{\prime}=0.128\ {\rm meV} \approx J/3, 
\eeq
\beq\label{D}
D=0.02\ {\rm meV},
\eeq
leading to a small ${\rm N{\acute e} el}$ 
temperature $T_N$ of less than $1K$ in the presence of the interlayer 
coupling $J^{\prime\prime}$.
The interlayer coupling is sufficiently small 
($J^{\prime\prime}/J = 0.045)\cite{coldea2}$ that we neglect 
it in our calculations performed at $T=0$.
The classical ground state for this Hamiltonian is a SDW whose 
wave vector $Q_0$ is along the
strong-coupling direction and has the value 
\beq\label{Q0}
Q_0 =2\pi (0.5+ \epsilon),
%\epsilon =(1/\pi)\arcsin{J^\prime/2J} 
\eeq
where experimentally, $\epsilon=0.030(2)$ is found 
for ${\rm Cs_2CuCl_4}$,\cite{coldea2} so that  $Q_0$ is close to $\pi$.

This means that individual chains are ordered approximately
antiferromagnetically, while the spins in adjacent chains, which are
slipped by half a lattice spacing along the chain direction, are
approximately orthogonal.

To further illustrate this point, consider the calculation
of the classical ground state energy of the SDW state, that can be easily
determined by introducing a local reference frame, so that at every 
site the averaged spin is directed along the $\zeta$ axis:
\beq\label{loc1}
S_{\vec R}^{x}=S_{\vec R}^{\xi} \cos (\vec Q \cdot \vec R)
-S_{\vec R}^{\zeta} \sin (\vec Q \cdot \vec R),\nonumber
\eeq
\beq
S_{\vec R}^{z}=S_{\vec R}^{\xi} \sin (\vec Q \cdot \vec R)
+S_{\vec R}^{\zeta} \cos (\vec Q \cdot \vec R),\nonumber
\eeq
\beq
S_{\vec R}^{y}=S_{\vec R}^{\eta}.
\eeq
In Eqs. (\ref{loc1}), $\vec Q$ is the wave vector 
of the spiral structure, which at the classical level we write as
$(Q_0, 0, 0)$ where $Q_0$ changes sign from layer to layer.
$Q_0$ is determined from minimization of the
classical ground state energy with $S_{\vec R}^{\xi}=S_{\vec R}^{\eta}=0$
for all $\vec R$. 
The classical ground state energy (per spin) is given by
\beq\label{EG0}
S^2 E_G^{(0)}(Q)=S^2 J^{T}_{\vec Q},
\eeq
\beq
J^{T}_{\vec Q}= J_{\vec Q}-iD_{\vec Q},
\eeq
where 
\beq\label{Jk}
 J_{\vec k}=J\cos{k_x}+2J^{\prime}\cos{\frac{k_x}{2}} 
\cos{\frac{\sqrt{3} k_y}{2}},
\eeq
and
\beq\label{Dk}
D_{\vec k}=-2iD \sin{\frac{k_x}{2}} 
\cos{\frac{\sqrt{3} k_y}{2}}
\eeq
are the Fourier transforms of the exchange and DM interactions of the
Hamiltonian respectively. 

For $D=0$, $\epsilon$ in Eq. (\ref{Q0}) is given by
\beq
\epsilon =\frac{1}{\pi}\arcsin \left (\frac{J^\prime}{2J} \right) = 0.0547,
\eeq
while for $D=0.02$ meV, the value of $Q_0$ that minimizes 
the classical ground state energy corresponds to $\epsilon =0.0533$
for the experimental values of $J$ and $J^\prime$.
The two terms in Eq. (\ref{Jk}), represent the intrachain and
interchain energies respectively. Substituting the actual value 
of $Q_0$ for a given $J^\prime /J$ and $D=0$, we find that
\beq
S^2 E_G^{(0)} (Q_0) = -JS^2\biggl[1+\frac{J'^2}{2J^2} \biggr].
\eeq
Thus the contribution to the ground state energy from the interchain
coupling is of order $(J^\prime)^2/J^2$ compared to the intrachain
contribution. Quantum fluctuations reduce the effectiveness of interchain
coupling further by renormalizing the SDW wave vector to a value even closer to $\pi$
(corresponding to a smaller effective $J^\prime /J$). At the same time,
quantum fluctuations lower the total energy.  Thus if one were to model
the effects of quantum fluctuations in terms of effective coupling
constants, $\tilde{J}$ and $\tilde{J}^\prime$, the result would be 
$\tilde{J}^\prime / \tilde{J} < J^\prime /J$ and $\tilde{J} > J$.
As is shown in the next section and in Appendix A, the value of $\epsilon$ 
obtained at the classical level, is in fact considerably renormalized 
by quantum fluctuations. 

\section{1/S Expansion}

To address the physics determined by quantum fluctuations, we
employ the well-known Holstein-Primakoff transformation for the 
spin operators:\cite{holstein}
\beq\label{holst1}
S_{\vec R}^{-}=S_{\vec R}^{\xi}-iS_{\vec R}^{\eta}\approx
\sqrt{2S} a^{\dagger}_{\vec R} \left [1- \frac{1}{4S} a^{\dagger}_{\vec R}
a_{\vec R} \right], 
\eeq
\beq\label{holst2}
S_{\vec R}^{+}=S_{\vec R}^{\xi}+iS_{\vec R}^{\eta}\approx
\sqrt{2S} \left [1- \frac{1}{4S} a^{\dagger}_{\vec R}
a_{\vec R} \right] a_{\vec R}, 
\eeq
\beq\label{holst3}
S_{\vec R}^{\zeta}= S- a^{\dagger}_{\vec R} a_{\vec R}.
\eeq
These are written  in a form with the square root expanded 
to first order in $1/(2S)$. The magnon operators 
$a^{\dagger}_{\vec R}$, $a_{\vec R}$ describe excitations around 
the spiral ground state and obey Bose statistics. Substituting the 
transformations
(\ref{holst1})-(\ref{holst3}) into Eq. (\ref{hamil}), we obtain the 
Hamiltonian for interacting magnons,
\beq\label{hamil1}
{\cal H}=S^2 E_G^{(0)}(Q)+({\cal H}^{(2)}+{\cal H}^{(3)}+
{\cal H}^{(4)}),
\eeq
where in the Fourier transformed representation
\beq\label{H2}
{\cal H}^{(2)}=2S\sum_{\vec k} \left( A_{\vec k} 
a^{\dagger}_{\vec k} a^{}_{\vec k}
-\frac{B_{\vec k}}{2} ( a^{\dagger}_{\vec k}  a^{\dagger}_{-\vec k} +
a^{}_{\vec k} a^{}_{-\vec k}) \right).
\eeq
The functions $A_{\vec k}$ and $B_{\vec k}$ are expressed through 
$J_{\vec k}$ as
\beq
A_{\vec k}=\frac14 \left[ J^T_{\vec Q+ \vec k}+J^T_{\vec Q- \vec k} \right] +
\frac{J_{\vec k}}{2} - J^T_{\vec Q},
\eeq
\beq
B_{\vec k}= \frac{J_{\vec k}}{2}
-\frac14 \left[J^T_{\vec Q + \vec k}+J^T_{\vec Q- \vec k} \right].
\eeq

The interactions between magnons are described by the last two
terms ${\cal H}^{(3)}$ and ${\cal H}^{(4)}$.
The three-magnon term can be written in the form
\beq\label{H3}
{\cal H}^{(3)}=\frac{i}{2} \sqrt{\frac{S}{2N}} \sum_{\vec 1,\vec 2,\vec 3} 
( C_{\vec 1} &+& C_{\vec 2} )
(a^{\dagger}_{\vec 3} a^{}_{\vec 2} a^{}_{\vec 1}\nonumber \\
&-& a^{\dagger}_{\vec 1} a^{\dagger}_{\vec 2} a^{}_{\vec 3}) 
\delta_{\vec 1+\vec 2,\vec 3}
\eeq
that contains the vertex
\beq
C_{\vec k}=\left( J^T_{\vec Q+ \vec k}-
  J^T_{\vec Q- \vec k} \right),
\eeq
which is antisymmetric with respect to the transformation 
$\vec k \rightarrow -\vec k$. For brevity, in longer expressions
we use the convention that $\vec 1 =\vec k_{1}$, etc.

The four-magnon term can be conveniently written as:
\begin{widetext}
\beq\label{H4}
\begin{array}{ll}
{\cal H}^{(4)}= \displaystyle\frac{1}{4N}\displaystyle
\sum_{\vec 1, \vec 2, \vec 3, \vec 4} &
\left\{ 
   
\left[ \left( A_{\vec 1-\vec 3}+A_{\vec 1-\vec 4} +A_{\vec
2-\vec 3}+A_{\vec 2-\vec 4} \right) - \left( B_{\vec 1-\vec 3}
+B_{\vec 1-\vec 4}+B_{\vec 2-\vec 3}+B_{\vec 2-\vec 4} \right) 
-\left( A_{\vec 1} +A_{\vec 2} +A_{\vec 3}+A_{\vec 4} \right)
\right]\vphantom{\displaystyle\frac23}\right.\\
&\left.
a^{\dagger}_{\vec 1} a^{\dagger}_{\vec 2}
a^{}_{\vec 3} a^{}_{\vec 4}
\delta_{\vec 1+\vec 2, \vec 3+\vec 4} 
 +\displaystyle\frac23
\left( B_{\vec 1} +
B_{\vec 2} +B_{\vec 3} \right) \left( a^{\dagger}_{\vec 1}
a^{\dagger}_{\vec 2} a^{\dagger}_{\vec 3} a^{}_{\vec 4}+ 
a^{\dagger}_{\vec 4} a^{}_{\vec 3} a^{}_{\vec 2}
a^{}_{\vec 1} \right) \delta_{\vec 1+\vec 2 +\vec 3, \vec 4} \right\} ,
\end{array}
\eeq
\end{widetext}
and contains only symmetric vertices. 
It is important to emphasize that  ${\cal H}^{(3)}$ comes from
the coupling between the operators $S_{\vec R}^{\xi}$ and
$S_{\vec R}^{\zeta}$,which arises for non-collinear ordered
states. This term plays
a crucial role in calculations of the structure factor as well as in
renormalization of the energy spectrum. 

To proceed, one needs to diagonalize the quadratic part of the 
Hamiltonian Eq. (\ref{H2}) by means of the Bogoliubov transformation:
\beq\label{bogol}
a^{}_{\vec k}=u_{\vec k}c^{}_{\vec k}+v_{\vec k}c^{\dagger}_{-\vec k},\quad
a^{\dagger}_{-\vec k}=u_{\vec k}c^{\dagger}_{-\vec k}+
v_{\vec k}c^{}_{\vec k},
\eeq
\beq\label{uv}
u_{\vec k}=\sqrt{\displaystyle\frac{A_{\vec k}+\varepsilon_{\vec k}}
{2\varepsilon_{\vec k}}}, \quad 
v_{\vec k}={\rm sgn} B_{\vec k}\sqrt{\displaystyle\frac{A_{\vec k}-
\varepsilon_{\vec k}}{2\varepsilon_{\vec k}}}.
\eeq
The energy spectrum in the above equations is given by:
\begin{eqnarray}
\label{spectrum}
\varepsilon_{\vec k}&=&\sqrt{A_{\vec k}^2 -B_{\vec k}^2} \nonumber \\
&=&\sqrt{(J_{\vec k}-J^T_{\vec Q})\left[(J^T_{\vec Q +\vec k}+
J^T_{\vec Q -\vec k})/2 - J^T_{\vec Q} \right] }.
\end{eqnarray}

For $D=0$, the magnon spectrum has zeros at $\vec k =0$ 
and $\vec k =\pm \vec Q$ in the two-dimensional Brillouin zone, 
while a non-zero $D$ leads to the appearance of a finite gap 
of order $\sqrt{DJ}$ around 
the $\vec Q$-points.
After diagonalization, ${\cal H}^{(2)}$ takes the form:
\beq\label{H2new}
{\cal H}^{(2)}=\displaystyle SE_G^{(1)}(Q) + 
\frac{2S}{N} \sum_{\vec k} \varepsilon_{\vec k} c^{\dagger}_{\vec k}c^{}_{\vec k},
\eeq
\beq\label{EG1}
E_G^{(1)}(Q) = J^T_{\vec Q} + \frac1N \sum_{\vec k} \varepsilon_{\vec k}.
\eeq

The contribution $S E_G^{(1)}(Q)$ gives the leading $1/(2S)$
correction to the classical ground state energy 
Eq. (\ref{EG0}). Minimizing the sum $S^2 E_G^{(0)}(Q)+SE_G^{(1)}(Q)$,
we find the quantum correction to the classical 
value $Q_0$.  To order $1/S$ the shifted wave vector is 
$\vec Q = (Q,0,0)$
\beq\label{qren}
Q=Q_0 +\displaystyle\frac{\Delta Q^{(1)}}{2S} ,
\eeq
\beq\label{Q1}
\Delta Q^{(1)}  = -\left[ 
\displaystyle\frac{\partial^2 J^T_{\vec Q}}{\partial Q^2}
\right] ^{-1} \frac1N \sum_{\vec k} 
\displaystyle\frac{( A_{\vec k}+B_{\vec k})}{\varepsilon_{\vec k}}
\cdot \frac{\partial J^T_{\vec Q +\vec k} }{\partial Q}
\Biggr\vert \sb{Q_0}.
\eeq
In the absence of the DM interaction,
\beq\label{Q1abs}
\Delta Q^{(1)}/2\pi = -0.0324,
\eeq
while for $D=0.02$ meV,
\beq\label{Q1with}
\Delta Q^{(1)}/2\pi= -0.0228.
\eeq
We see that inclusion of the Dzyaloshinskii-Moriya interaction
suppresses the renormalization of $Q$ towards $\pi$.

It is possible to go further and also calculate the correction 
to the ground state energy and the ordering wave vector $Q$ that is
of order $1/(2S)^2$. The details of those calculations are presented
in Appendix A. The results indicate that quantum corrections
computed order by order in $1/(2S)$ are significant for the values of
$J$ and $J^{\prime}$ given by Eqs. (\ref{J}).
To further explore this issue, one can compare the classical
ground state energy Eq. (\ref{EG0}) to its leading
renormalization due to quantum fluctuations $S E_G^{(1)}(Q_0)$.
We find, for $D=0$
\begin{eqnarray}\label{ge0} 
S^2 E_G^{(0)}(Q_0)/J=-0.265, \ \ S E_G^{(1)}(Q_0)/J=-0.157,
\end{eqnarray}
and, for $D=0.02$ meV,
\begin{eqnarray}
S^2 E_G^{(0)}(Q_0)/J=-0.291, \quad S E_G^{(1)}(Q_0)/J=-0.138.
\end{eqnarray}
We see that the leading quantum correction 
lowers the ground state energy by about 50\%, giving reasonable 
agreement with a recent experimental determination of this quantity, 
based on susceptibility measurements,\cite{tokiwacold} which yields 
a ground state energy slightly above $-0.5J$. 
We show in Appendix A that for $D=0$ the next order in $1/(2S)$ 
correction  reduces the ground state energy by a further $10$ 
percent below the sum of the values given in Eq. (\ref{ge0}), 
bringing the total within a few percent of the experimental value. 
 
To calculate physical observables, we need the Green's functions
for the magnon operators $a_{\vec k}$. They can be
conveniently written in the form of a
$2\times 2$ matrix, indicating the presence of normal 
and anomalous parts. At zero temperature, the definition reads
\begin{eqnarray}
i{\hat G} (\vec k, \omega)=  \int_{-\infty}^{\infty} d t e^{i \omega t}
 \bigg\langle {\hat T} \left[
\begin{tabular}{l}
$a^{}_{\vec k} (t)$  \\ $a^{\dagger}_{-\vec k} (t)$
\end{tabular} \right]
\left[ a^{\dagger}_{\vec k}(0) a^{}_{-\vec k }(0) 
\right] \bigg\rangle , \nonumber\\
\end{eqnarray}
where ${\hat T}$ is the time ordering operator, and the average is
taken with respect to the ground state.
The inverse of the matrix for the bare Green's functions has the form
\cite{ohyama}
\begin{widetext}
\beq\label{Ginv0} 
{\hat G}^{(0) -1} (\vec k, \omega)=
 \left( \begin{array}{cc}
\omega -2SA_{\vec k}+i\eta & 2SB_{\vec k}\\
 2SB_{\vec k} & -\omega -2SA_{\vec k}+i\eta 
\end{array} \right), 
\eeq
corresponding to the Green's function
\beq\label{G0} 
{\hat G}^{(0)} (\vec k, \omega)=
 \displaystyle\frac{1}{(\omega-\omega_{\vec k}+i\eta)
(\omega+\omega_{\vec k}-i\eta)}
\left( \begin{array}{cc}
\omega +2SA_{\vec k}-i\eta & 2SB_{\vec k}\\
 2SB_{\vec k} & -\omega +2SA_{\vec k}-i\eta 
\end{array} \right), 
\eeq
where $\omega_{\vec k}=2S\varepsilon_{\vec k}$.
The self-energy  ${\hat \Sigma}(\vec k, \omega)$ determining the exact
Green's function is also a $2\times 2$ matrix, that
can be parametrized as
\beq 
{\hat \Sigma} (\vec k, \omega)=
 \left( \begin{array}{cc}
X(\vec k, \omega)+Y(\vec k, \omega) & Z(\vec k, \omega) \\
 Z(\vec k, \omega) & X(\vec k, \omega)-Y(\vec k, \omega)
\end{array} \right), 
\eeq
\end{widetext}
and which satisfies the Dyson equation
\beq\label{Dys}
{\hat G}^{-1}(\vec k, \omega)={\hat G}^{(0) -1}(\vec k, \omega)
-{\hat \Sigma}(\vec k, \omega).
\eeq

The self-energy to order $1/(2S)$ consists of two parts
\beq
{\hat \Sigma}(\vec k, \omega)={\hat \Sigma}^{(4)}(\vec k)
+{\hat \Sigma}^{(3)}(\vec k, \omega).
\eeq
The contribution ${\hat \Sigma}^{(4)}(\vec k)$ is frequency 
independent and can be calculated simply by the Hartree-Fock 
decoupling of the quartic term in the Hamiltonian Eq. (\ref{H4}).
\begin{eqnarray}\label{sigX4}
X^{(4)}(\vec k)&=&A_{{\vec k}} +\frac{2 S}{N} \sum_{{\vec
q}} \frac{1}{\omega_{\vec q}} \bigg
[ \left(\frac{1}{2} B_{{\vec k}} +B_{\vec q} \right) B_{\vec
q} \nonumber \\ &+& \left( A_{{\vec k}-{\vec q}}
-B_{{\vec k}-{\vec q}} -A_{\vec q} -A_{\vec k} \right)
A_{\vec q} \bigg],
\end{eqnarray}
\begin{eqnarray}\label{sigY4}
Y^{(4)}(\vec k)&=&-B_{{\vec k}} +\frac{2 S}{N} \sum_{{\vec q}}
\frac{1}{\omega_{\vec q}} \bigg[ \left( B_{{\vec k}} +B_{\vec
q} \right) A_{\vec q} \nonumber \\ &+& \left(A_{{\vec
k}-{\vec q}} -B_{{\vec k}-{\vec q}} -A_{\vec q}
- \frac{1}{2} A_{{\vec k}} \right) B_{\vec q} \bigg], 
\label{Self2}
\end{eqnarray}
\beq\label{sigZ4}
Z^{(4)}(\vec k)=0. 
\eeq
${\hat \Sigma}^{(3)}(\vec k, \omega)$ is the contribution
arising from the three-magnon interactions, ${\cal H}^{(3)}$. Its
non-zero imaginary part determines the magnon damping to first 
order in $1/(2S)$. The corresponding components are most easily 
calculated by transforming Eq. (\ref{H3}) to the 
$c_{\vec k}$-operator basis using the Bogoliubov 
transformations, Eqs. (\ref{bogol}).
\begin{widetext}
\beq\label{sigX3}
X^{(3)} (\vec k,\omega)& =& -\frac{S}{16N} \sum_{\vec q}
\left\{ \left[ \Phi_1 ( \vec q, \vec k-\vec
q)\right]^2 +\left[ \Phi_2 ( \vec q, \vec k
-\vec q) \right]^2 \right\} \left( 
\frac{1}{ \omega_{\vec q} + \omega_{\vec k-\vec q} 
- \omega - i\eta } +
\frac{1}{ \omega_{\vec q} + \omega_{\vec k-\vec q}
+ \omega - i\eta } \right),\nonumber
\eeq
\beq\label{sigZ3}
Z^{(3)}(\vec k, \omega) &=
& -\frac{S}{16N} \sum_{\vec q} \left\{ \left[ \Phi_1
( \vec q, \vec k-\vec q)\right]^2 -\left
[ \Phi_2 ( \vec q, \vec k-\vec q) \right]^2
\right\} \left( \frac{1}{ \omega_{\vec q} + \omega_{\vec
k-\vec q} - \omega - i \eta}+ \frac{1}{ \omega_{\vec
q} + \omega_{{\vec k}-{\vec q}} + \omega - i \eta}
\right),\nonumber 
\eeq
\beq\label{sigY3} 
Y^{(3)}(\vec k, \omega) &=& -\frac{S}{16N}
\sum_{{\vec q}} \left\{ 2 \Phi_{1}( {\vec q}, {\vec
k}-{\vec q}) \Phi_{2}( {\vec q}, {\vec k}-{\vec
q})\right\} \left( \frac{1}{ \omega_{\vec q} +
\omega_{{\vec k}-{\vec q}} - \omega - i \eta}- \frac{1}
{ \omega_{\vec q} + \omega_{{\vec k}-{\vec q}} + \omega
- i \eta} \right).
\eeq
\begin{eqnarray}\label{phi}
\Phi_1({\vec q}, {\vec k}-{\vec q}) &=& \left
( C_{\vec q} +C_{{\vec k}-{\vec q}} \right) \left
( u_{\vec q} +v_{\vec q} \right) \left( u_{{\vec
k}-{\vec q}} +v_{{\vec k}-{\vec q}} \right) -2 C_{{\vec
k}} \left( u_{{\vec q}} v_{{\vec k}-{\vec q}} +v_{{\vec
q}} u_{{\vec k}-{\vec q}} \right), \nonumber \\
\Phi_2 ({\vec q}, {\vec k}-{\vec q}) &=& C_{\vec
q} \left( u_{\vec q} +v_{\vec q} \right)
\left( u_{{\vec k}-{\vec q}} -v_{{\vec k}-{\vec q}}
\right)+C_{{\vec k}-{\vec q}} \left( u_{{\vec k}-{\vec
q}} +v_{{\vec k}-{\vec q}} \right) \left( u_{{\vec
q}} -v_{{\vec q}} \right).
\end{eqnarray}
\end{widetext}

\section{Sublattice Magnetization}

In this section we consider the staggered magnetization 
for a range of values of $J^{\prime}/J$ and $D/J$ and show
that quantum fluctuations lead to a considerable renormalization
of the average value of the local moment.
From the operator definition Eq.~(\ref{holst3})
we can write the staggered magnetization as
\beq\label{stmdef}
M=S+\frac1N \sum_{\vec k}\int \frac{d\omega}{2\pi i}
 G_{11}(\vec k, \omega)e^{-i\omega 0_{-}},
\eeq
where the exact Green's function $G_{11}(\vec k, \omega)$ is the 
corresponding element of the full matrix ${\hat G}(\vec k, \omega)$ 
and $0_{-}$ is a negative infinitesimal. To lowest order in
$1/(2S)$, one can use  $G_{11}^{(0)}(\vec k, \omega)$ of
Eq.~(\ref{G0}), to find the first order correction to the $S=1/2$ value
\beq\label{stm1}
\Delta M^{(1)}=
-\displaystyle\frac{1}{2N} \sum_{\vec k} \left(  
\displaystyle\frac{A_{\vec k}}{\varepsilon_{\vec k}} -1 \right).
\eeq
For the next order correction in $1/(2S)$, 
there are two contributions.
The first one, denoted as $\Delta M^{(2)}_{{\rm I}}$,
arises from renormalization by quantum fluctuations of the
ordering wave vector $Q_0$. This correction is obtainable by 
substituting the renormalized value $Q$ given by Eq.~(\ref{qren}) into the 
formula above and expanding using Eq. (\ref{stm1}) up to first 
order in $1/(2S)$.
\beq\label{M2I}  
\Delta M^{(2)}_{{\rm I}}=\displaystyle\frac{\Delta Q^{(1)}}{2S}
\frac1N \sum_{\vec k} 
\displaystyle\frac{B_{\vec k}( A_{\vec k}+B_{\vec k})}{4\varepsilon_{\vec k}^3}
\cdot \frac{\partial J^T_{\vec Q +\vec k}}{\partial Q}
\Biggr\vert \sb{Q_0}.
\eeq
The second contribution is determined by the self-energy corrections
of order $1/(2S)$ to the Green's function itself and is given by
\begin{widetext}
\beq\label{M2II}
\Delta M^{(2)}_{{\rm II}} = \frac1N 
\sum_{\vec k}\int \frac{d\omega}{2\pi i}
e^{-i\omega 0_{-}} \Delta {\hat G}_{11}(\vec k, \omega),
\eeq
where
\begin{eqnarray}\label{delG11}
&\Delta G_{11}(\vec k, \omega)=
\left( {\hat G}^{(0)}(\vec k, \omega){\hat \Sigma}(\vec k, \omega)
 {\hat G}^{(0)}(\vec k, \omega) \right)_{11}=&\nonumber\\
&\left[ X(\vec k, \omega)+Y(\vec k, \omega)\right]
\left[ G^{(0)}_{11}(\vec k, \omega)\right]^2
+2Z(\vec k, \omega) G^{(0)}_{11}(\vec k, \omega)
G^{(0)}_{12}(\vec k, \omega)
+\left[ X(\vec k, \omega)-Y(\vec k, \omega)\right]
\left[ G^{(0)}_{12}(\vec k, \omega)\right]^2.&
\end{eqnarray}
This correction, as a result of integration over $\omega$, 
will contain two parts, so that
\beq 
\Delta M^{(2)}_{{\rm II}}= \Delta M^{(2)}_{a}+ \Delta M^{(2)}_{b}.
\eeq 
One part comes from taking the residues at the poles of self-energies, 
while the other results from the double-poles of the 
products of two Green's functions. Using
Eqs. (\ref{sigX4})-(\ref{sigY3}), we obtain the first contribution
\begin{eqnarray}\label{stma}
&\Delta M^{(2)}_{a}   =-\displaystyle\frac{S}{8N} \sum_{\vec k,\vec q}
\left\{ 
\displaystyle\frac{\left[ \Phi_1 ({\vec q}, {\vec k}-{\vec q}) \right]^2}{2}
\left[
G^{(0)}_{11}(\vec k, \Omega_{\vec k,\vec q})
+G^{(0)}_{12}(\vec k, \Omega_{\vec k,\vec q})
\right]^2 
+\displaystyle\frac{ 
\left[ \Phi_2 ({\vec q}, {\vec k}-{\vec q}) \right]^2 }{2}\cdot 
\right.\nonumber\\
&\left. \left[ G^{(0)}_{11}(\vec k, \Omega_{\vec k,\vec q})
-G^{(0)}_{12}(\vec k, \Omega_{\vec k,\vec q})
\right]^2 
 -\Phi_1 ({\vec q}, {\vec k}-{\vec q})  
\Phi_2 ({\vec q}, {\vec k}-{\vec q})
\left[ \left[ G^{(0)}_{11}
(\vec k, \Omega_{\vec k,\vec q})\right]^2
 -\left[ G^{(0)}_{12}(\vec k, \Omega_{\vec k,\vec q}) \right]^2 \right] 
\right\},
\end{eqnarray}
\beq\label{omkq}
\Omega_{\vec k,\vec q}=-\omega_{\vec q}-\omega_{\vec k-\vec q}.
\eeq
The contribution arising from the residues at the
double-poles of the products of two Green's functions in 
Eqs. (\ref{M2II})-(\ref{delG11}), is straightforward to calculate as
well. After some algebra, we find that
\begin{eqnarray}\label{stmb}
&\Delta M^{(2)}_{b} = \displaystyle\frac{1}{2SN} \sum_{\vec k}
\displaystyle\frac{B_{\vec k}}{4\varepsilon_{\vec k}^3}
\bigg\{ \left[ X(\vec k, \omega_{\vec k})+Z(\vec k, \omega_{\vec k})
\right]  \left[ A_{\vec k}+B_{\vec k}\right]-
\left[ X(\vec k, \omega_{\vec k})-Z(\vec k, \omega_{\vec k})
\right]  \left[ A_{\vec k}-B_{\vec k} \right] \bigg\} \nonumber\\
&+\displaystyle\frac1N \sum_{\vec k} 
\displaystyle\frac{(A_{\vec k}-\varepsilon_{\vec k})}
{4\varepsilon_{\vec k}^2}
\bigg\{  \left[ X^{\prime}(\vec k, \omega_{\vec k})
+Z^{\prime}(\vec k, \omega_{\vec k})
\right]  \left[ A_{\vec k}+B_{\vec k}\right]-
\left[ X^{\prime}(\vec k, \omega_{\vec k})-Z^{\prime}(\vec k, \omega_{\vec k})
\right]  \left[ A_{\vec k}-B_{\vec k} \right]
-2 Y^{\prime}(\vec k, \omega_{\vec k})\varepsilon_{\vec k}
\bigg\},
\end{eqnarray}
where $\omega_{\vec k}=2S\varepsilon_{\vec k}$, and
\beq 
\bigg\{ X^{\prime}(\vec k, \omega_{\vec k}),
Y^{\prime}(\vec k, \omega_{\vec k}),
Z^{\prime}(\vec k, \omega_{\vec k}) \bigg\}=
\bigg\{ 
\frac{\partial X(\vec k, \omega)}{\partial \omega},
\frac{\partial Y(\vec k, \omega)}{\partial \omega},
\frac{\partial Z(\vec k, \omega)}{\partial \omega} \bigg\}
\bigg\vert\sb{\omega=\omega_{\vec k}}.
\eeq
\end{widetext}
It is clear from the expressions above that both 
$\Delta M^{(2)}_{{\rm I}}$ 
and $\Delta M^{(2)}_{{\rm II}}$ contain the overall 
prefactor $1/(2S)$.

\begin{figure}
\includegraphics[height=6cm, width=8cm]{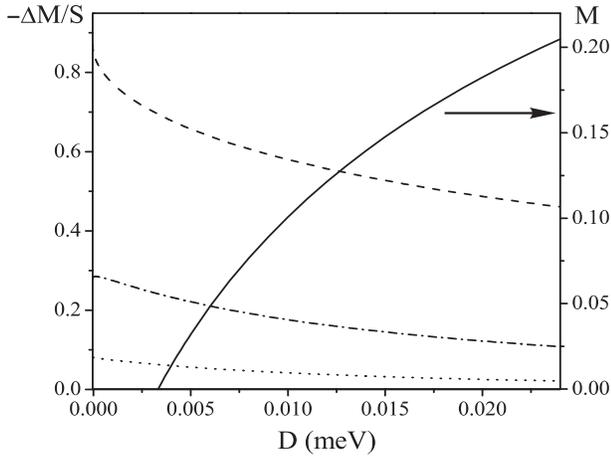}
\caption{The total value of sublattice magnetization $M$, 
Eq. (\ref{magtot}), calculated up to the second order in $1/(2S)$  
as a function of Dzyaloshinskii-Moriya interaction $D$, is shown by the
solid line. The dashed, dotted and dashed-dotted lines are the results
for the quantum 
corrections $-\Delta M^{(1)}$,
$-\Delta M^{(2)}_{{\rm I}}$ and $-\Delta M^{(2)}_{{\rm II}}$,
respectively, divided by $S=1/2$.}
\label{mag1}
\end{figure}

In Fig. ~\ref{mag1}, we plot the total sublattice magnetization 
\beq\label{magtot}
M=S+\Delta M^{(1)}+\Delta M^{(2)}_{{\rm I}}+\Delta M^{(2)}_{{\rm II}}
\eeq
as a function of Dzyaloshinskii-Moriya interaction $D$, for coupling
constants $J$ and $J^{\prime}$ as given by Eqs. (\ref{J}). 
The ratios of each correction 
to the classical magnetization $S=1/2$, $-\Delta M/S$,
are also plotted, as explained in the caption of Fig. ~\ref{mag1}. 
One can see that the renormalization of the local 
magnetization by quantum fluctuations to second order in $1/(2S)$ is
considerable for all values of $D$ between $0$ and $0.025$ meV, although 
increasing $D$ clearly suppresses fluctuations making the system 
more classical. The  figure also shows a calculated sublattice 
magnetization that is negative for $D$ between $0$ and 
$D\approx 0.003$ meV, signaling the melting of the assumed
spiral ordered ground state by quantum fluctuations when $D$ is too small.

Experimentally, a value of $-\Delta M/S \approx 0.25$ was 
inferred by Coldea and 
co-workers\cite{coldea3} from elastic neutron scattering, and this value 
was used in the analysis of 
the most recent neutron data.\cite{coldea} Such a small 
renormalization of the local moment is incompatible with the 
theory presented here. It is also unlikely to be
explained by the addition of the very small expected interlayer exchange 
coupling $J^{\prime\prime}/J=0.045$.\cite{coldea, coldea2} It is indeed 
puzzling that such a small moment reduction, which is less than 
for the spin $1/2$ Heisenberg antiferromagnet on a two-dimensional
square lattice, would occur in this system which is both frustrated and
close to the one-dimensional limit. 

\section{Energy Spectrum}
The renormalized magnon energy spectrum 
${\tilde \omega}_{\vec k}=2S{\tilde \varepsilon}_{\vec k}$
can be found from the poles of the exact Green's function, 
determined by\cite{ohyama}
\begin{widetext}
\begin{eqnarray}
{\rm Re} \left\{ {\rm det}{\hat G}^{-1}(\vec k, {\tilde \omega}_{\vec k})
\right\} =
{\rm Re}\left\{ {\rm det}\left[ 
{\hat G}^{(0) -1}(\vec k, {\tilde \omega}_{\vec k})
-{\hat \Sigma}(\vec k, {\tilde \omega}_{\vec k}) \right] \right\}=0,
\end{eqnarray}
which leads to the self-consistency equation:
\begin{eqnarray}\label{disp}
{\tilde \omega}_{\vec k}=&
\left\{  
\left\{ 2S ( A_{\vec k}+B_{\vec k})+
{\rm Re} \left[ X(\vec k, {\tilde \omega}_{\vec k})-
Z(\vec k, {\tilde \omega}_{\vec k}) \right] \right\}
\left\{ 2S ( A_{\vec k}-B_{\vec k})+
{\rm Re} \left[ X(\vec k, {\tilde \omega}_{\vec k})+
Z(\vec k, {\tilde \omega}_{\vec k}) \right] \right\} \right.\nonumber\\
&\left. -\left\{ {\rm Im} \left[ 
X(\vec k, {\tilde \omega}_{\vec k})+
Z(\vec k, {\tilde \omega}_{\vec k}) \right] \right\}
\left\{ {\rm Im} \left[  
X(\vec k, {\tilde \omega}_{\vec k})-
Z(\vec k, {\tilde \omega}_{\vec k}) \right] \right\}
+\left[ {\rm Im} Y(\vec k, {\tilde \omega}_{\vec k}) \right]^2 
\right\}^{1/2} +{\rm Re} Y(\vec k, {\tilde \omega}_{\vec k}).
\end{eqnarray}
Near these poles, if the imaginary part of the self-energy is small,
the shape of the one-magnon spectrum is approximately  
Lorentzian and has the form\cite{ohyama}

\beq\label{lor}
{\rm det} {\hat G}^{-1} (\vec k, \omega) \simeq
-2\omega_{\vec k} z_{\vec k} (\omega-{\tilde \omega}_{\vec k})
+i{\rm Im}\left[ {\rm det} {\hat G}^{-1} 
(\vec k,{\tilde \omega}_{\vec k} ) \right],
\eeq
\end{widetext}
where 
\beq\label{omz}
2\omega_{\vec k } z_{\vec k}=-\frac{\partial}{\partial\omega}
{\rm Re} \left[ {\rm det} {\hat G}^{-1} 
(\vec k, \omega ) \right] \bigg\vert_{\omega ={\tilde
\omega_{\vec k} } }.
\eeq 
The imaginary part of Eq. (\ref{lor}) determines the inverse life-time
of the quasiparticles. To calculate the half-width at half-maximum of the
peak, we use the parameter 
\beq\label{gamma}
\Gamma_{\vec k}\simeq \frac{\left| {\rm Im} [ {\rm det} {\hat G}^{-1} 
(\vec k, {\tilde \omega}_{\vec k} ) ] \right| }
{2\omega_{\vec k } z_{\vec k}}.
\eeq

Equation (\ref{disp}), although formally exact, implicitly includes
contributions of all orders in $1/S$.  Furthermore, quantities such 
as the terms 
$2S ( A_{\vec k} \pm B_{\vec k})$ that are of the leading order in $S$,
should be corrected for the first order shift of the ordering wave vector 
from $\vec Q_0$ to $\vec Q$, given by Eqs. (\ref{qren})-(\ref{Q1}). Although
the self-energies are only calculated to order $1/S$, taking the square root 
of a sum of squares of self-energies evaluated at renormalized frequencies, 
effectively mixes in higher order corrections. Also,
particularly for the case $D=0$, there is cancellation among terms of order
$1/S$, leaving a result, at the shifted $Q$-vector, 
which is the square root of 
residual $1/S^2$ terms, which may be negative.  

A simple way to preserve physical behavior at the Goldstone wave 
vector $\vec Q$ for $D=0$, is
to evaluate the self-energies using renormalized coupling constants, 
for which the classical ordering wave vector is $\vec Q$, that is  
\beq\label{Qn}
Q=\pi + 2\arcsin \left (\frac{{\bar J}^{\prime}}{2{\bar J}} \right).
\eeq
The effect of this renormalization, which determines the ratio 
${\bar J}^{\prime}/{\bar J}$ and represents a higher order in $1/S$ 
correction to the self-energies, is that the self-energies will vanish
at the renormalized $\vec Q$.  A further 
renormalization of ${\bar J}$ by itself is defined by the condition that 
the coefficient of the linear term in $|{\vec k}-{\vec Q}|$ inside the 
square root vanishes for ${\vec k}={\vec Q}$, so that the energy dispersion 
relation around this point is linear in $|{\vec k}-{\vec Q}|$.
The procedure of 
expanding the self-energies around $\vec Q$ is tedious but completely 
equivalent to that described elsewhere.\cite{ohyama,chubuk} 
Omitting the 
intermediate steps, we obtain the condition
\beq\label{cons}
\frac{\partial J_{\vec k}}{\partial \vec k} \bigg\vert_{\vec k =\vec Q}
=-\frac{1}{2SN}\sum_{\vec p} 
\frac{ ({\bar A}_{\vec p}+{\bar B}_{\vec p}) }
{{\bar \varepsilon}_{\vec p}} \cdot
\frac{\partial {\bar J}_{\vec p+\vec Q}}{\partial \vec p},
\eeq
where the bars indicate that the renormalized values ${\bar J}$ 
and ${\bar J}^{\prime}$ should be used in the right hand side 
of this equation. One can easily check that, within this approach, we 
retain consistency up to lowest order in $1/2S$. 
Indeed, expanding the derivative $\partial J_{\vec k}/\partial \vec k$ 
around $Q_0$, we recover Eq. (\ref{Q1}) for the correction to the 
ordering wave vector. Using Eqs. (\ref{J}), we find
from Eqs. (\ref{Qn})-(\ref{cons}) that 
\beq\label{barJLp}
{\bar J}\approx 0.472\ {\rm meV},\quad  {\bar J}^{\prime}
\approx 0.066\ {\rm meV}.
\eeq
We see that in this approach, $J^{\prime}$ is 
renormalized down from its bare value, while $J$ increases, as 
discussed in Sec. II.

\begin{figure}
\includegraphics[width=\columnwidth]{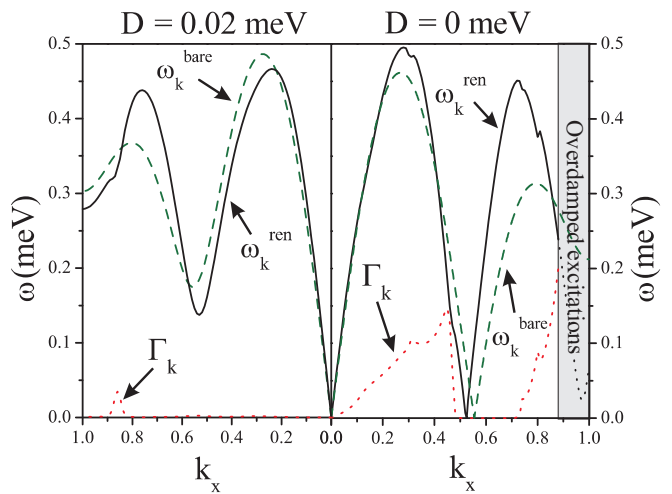}
\caption{(Color Online) The two panels present the results for the bare and 
renormalized energy spectra for the cases of $D=0$ and $D=0.02$ meV.
Spin-wave energies are shown together with
the inverse lifetime $\Gamma_{\vec k}$. The path in the
two-dimensional Brillouin zone sweeps from (00) point to (10). 
$k_x$, $k_y$ are measured in units of $2\pi$ and $2\pi/\sqrt{3}$
respectively. 
} 
\label{dispJo}
\end{figure}

In spite of the fact that we expect the system to be unstable 
for $D=0$ and $J^{\prime}/J = 1/3$, it is of pedagogical interest
to examine the effects of anharmonicity and the shift of the
Goldstone mode for this case.  The right hand 
panel of Fig.~\ref{dispJo} 
shows the results for the
renormalized energy spectrum and its comparison to the linear
spin-wave dispersion along the $(1,0)$ direction for $D=0$. 
The inverse lifetime of quasiparticles Eq.~(\ref{gamma}) 
is also shown. 
The excitations are seen to be heavily overdamped close to 
the $(10)$ point. 
Also kinks are seen at various places in the renormalized curves,
apparently at points where one-to-two magnon decay channels turn on.
Although not shown in Fig.~\ref{dispJo}, we also find that the band 
width of the dispersion along $(1,0) \rightarrow (1,1)$ is 
renormalized downward by about a factor of 3 due to anharmonic effects. 

%\begin{figure}
%\includegraphics[width=\columnwidth]{disp_both_dir.eps}
%\caption{The renormalized energy spectrum
%$\omega_{\vec k}^{{\rm ren}}$ (solid line), and the bare spectrum 
%$\omega_{\vec k}^{{\rm bare}}$ (dashed line) are shown together with
%the inverse lifetime $\Gamma_{\vec k}$ (dotted line). The path in the
%two-dimensional Brillouin zone sweeps from (00) point to (10), 
%and then further to (11).
%$k_x$, $k_y$ are measured in units of $2\pi$ and $2\pi/\sqrt{3}$
%respectively. 
%}\label{dispbd}
%\end{figure}

In the left hand panel of Fig.~\ref{dispJo}, 
we plot the  
renormalized energy spectrum for $D=0.02$ meV.\cite{veillette} 
In the presence of this DM 
interaction, it suffices to use the bare values for the coupling 
constants in expressions for self-energies, since the finite gap in the 
spectrum does not lead to any problematic behavior near $\vec Q$ 
when $D$ is not too small.
The effects of quantum fluctuations are very sensitive 
to the presence of a non-zero $D$. Increasing the value of $D$ 
reduces the interaction-induced damping of quasiparticles, making them 
more stable. Quantum fluctuations, in addition to shifting $Q$ 
towards $\pi$, reduce the size of the gap at this point.
However, the effect of suppression of the gap towards zero, is
relatively small for $D=0.02$ meV.

\section{Scattering Intensity}
In this Section, the dynamical properties of an antiferromagnet 
described by the model of Eq. (\ref{hamil}) are calculated within the 
framework of the $1/(2S)$ expansion. We are specifically interested 
in answering the question of whether anharmonic spin waves can account for
the main features seen in inelastic neutron scattering measurements 
on ${\rm Cs_2 Cu Cl_4}$.\cite{coldea}
We begin with expressions for the structure
factor components that enter the formula for the inelastic 
neutron scattering cross section. These results were obtained earlier 
by Ohyama and Shiba,\cite{ohyama} and we simply quote them here. 

Neutron scattering spectra of spin excitations in magnetic solids may be 
expressed in terms of the Fourier transformed real-time 
dynamical correlation function, $i,k=(x,y,z)$
\beq
S^{ik}(\vec k, \omega)= \displaystyle\frac{1}{2\pi}
\int_{-\infty}^{\infty}
dt \sum_{\vec R} \langle S_{\vec 0}^{i} (0) S_{\vec R}^{k} (t) \rangle
e^{i(\omega t -\vec k \cdot \vec R)}.
\eeq
For the spiral spin density wave state,  
$S^{xx}(\vec k, \omega)=S^{zz}(\vec k, \omega)$. The complete
expression for the inelastic, differential scattering
cross-section, including polarization factors, is 
given by\cite{lovesey}
\beq\label{totint}
I(\vec k, \omega)= p_x S^{xx}(\vec k, \omega)+ p_y S^{yy}(\vec k, \omega),
\eeq
\beq
p_x=1+\cos^2 \alpha_{\vec k}, \qquad p_y= \sin^2 \alpha_{\vec k},
\eeq
where $\alpha_{\vec k}$ is the angle between the scattering wave vector 
$\vec k$ and the axis perpendicular to the plane of the spins. 

We begin by calculating the corresponding
time-ordered spin-spin correlation functions in the rotated 
coordinate system
\beq
iF^{\alpha \beta}({\vec k}, \omega)= \int_{-\infty}^{\infty} d t
e^{- i \omega t } \langle {\hat T} S^{\alpha}_{-\vec k} (0) S^{\beta}_
{\vec k}(t) \rangle ,
\eeq
where $\alpha, \beta=(\xi, \eta, \zeta)$ are the rotated coordinate axes
given by Eqs. (\ref{loc1}).
The dynamical structure factor is related to the imaginary parts of 
these correlation functions in the following way:
\begin{widetext}
\begin{eqnarray}
S^{yy}({\vec k}, \omega)&=& -\frac{1}{\pi} {\rm Im} F^{\eta \eta}({\vec k},
\omega), \label{Syy} \\ 
S^{xx}({\vec k}, \omega)&=& S^{zz}({\vec k}, \omega)
= -\frac{1}{\pi} {\rm Im} \left[ \Theta^{+}({\vec k}+{\vec Q},
\omega) + \Theta^{-}({\vec k}-{\vec Q}, \omega) \right], \label{Sxx}\\
S^{xz}({\vec k}, \omega)&=& -S^{zx}({\vec k}, \omega)= -\frac{i}{\pi}
{\rm Im} \left [ \Theta^{+}( {\vec k}+{\vec Q}, \omega) -
\Theta^{-}({\vec k}-{\vec Q}, \omega) \label{Sxz} \right],
\end{eqnarray}
with,
\begin{equation}
\Theta^{\pm}({\vec k}, \omega)= \frac{1}{4} \left\{ F^{\xi \xi}({\vec k},
\omega) +F^{\zeta \zeta}({\vec k}, \omega) \mp i \left[ F^{\xi \zeta}
({\vec k},\omega) - F^{\zeta \xi}({\vec k}, \omega) \right] \right\}.
\end{equation}

We treat the interactions between the magnons as a perturbation
in the formally small parameter $1/(2S)$, and use the 
standard ${\hat S}$-matrix expansion while calculating the averages of 
the magnon operators $a_{\vec k}(t)$.\cite{abrikos}
As a result,
\begin{eqnarray}\label{1M}
F^{\eta \eta}({\vec k}, \omega)&=& \frac{S}{2} c^2_{y}
\left[ G_{11}({\vec k}, \omega)+ G_{22}({\vec k}, \omega)
-2 G_{12}({\vec k}, \omega) \right], \nonumber \\
F^{\xi \xi}({\vec k}, \omega)&=& \frac{S}{2} c^2_{x} 
\left[ G_{11}({\vec k}, \omega)+ G_{22}({\vec k}, \omega)
+2 G_{12}({\vec k}, \omega) \right],
\end{eqnarray}
where the Green's function are determined from Eq. (\ref{Dys}) 
with the self-energies given by Eqs. (\ref{sigX4})-(\ref{sigY3}), and,
to the relevant order in $1/(2S)$,
\begin{eqnarray}
c_{y}&=& 1- \frac{1}{4SN} \sum_{{\vec k}} \left( 2 v^2_{{\vec k}}-
u_{{\vec k}} v_{{\vec k}} \right), \nonumber \\ 
c_{x}&=& 1- \frac{1}{4SN} \sum_{{\vec k}} 
\left( 2 v^2_{{\vec k}}+ u_{{\vec k}} v_{{\vec k}}
\right).
\label{cxy}
\end{eqnarray}
The term that mixes the transverse and longitudinal fluctuations 
has the form:
\begin{eqnarray}
i\left[ F^{\xi \zeta}({\vec k}, \omega) -F^{\zeta \xi}
({\vec k}, \omega) \right] =& c_{x} 
\left\{ P^{(1)}({\vec k}, \omega) 
\left[ G_{11}({\vec k}, \omega)+ G_{22}({\vec k}, \omega)
-2 G_{12}({\vec k}, \omega) \right]\right.\nonumber\\
&\left.+  P^{(2)}({\vec k},\omega) \left[ G_{11}({\vec k}, \omega)- 
G_{22}({\vec k}, \omega) \right] 
 \right\}.
\end{eqnarray}
with the functions $P^{(1, 2)}({\vec k}, \omega)$ defined as
\begin{eqnarray}
P^{(1)}({\vec k}, \omega)= &\displaystyle\frac{S}{4N}& 
\sum_{{\vec q}}
\Phi^{(1)} \left( {\vec q}, {\vec k}-{\vec q} \right)
\left( u_{{\vec q}} v_{{\vec k} -{\vec q}} + v_{{\vec
q}} u_{{\vec k}-{\vec q}} \right)
\left( \frac{1}{ \omega_{\vec q} + \omega_{{\vec k}-{\vec
q}} - \omega - i \eta}+ \frac{1}{ \omega_{\vec q} +
\omega_{{\vec k}-{\vec q}} + \omega - i \eta} \right),
\nonumber \\ 
P^{(2)}({\vec k}, \omega) = &\displaystyle\frac{S}{4N}& \sum_{{\vec
q}} \Phi^{(2)} \left( {\vec q}, {\vec k}-{\vec
q} \right) \left( u_{{\vec q}} v_{{\vec k} -{\vec
q}} + v_{{\vec q}} u_{{\vec k}-{\vec q}}
\right)
\left( \frac{1}{ \omega_{\vec q} + \omega_{{\vec k}-{\vec
q}} - \omega - i \eta}- \frac{1}{ \omega_{\vec q} +
\omega_{{\vec k}-{\vec q}} + \omega - i \eta} \right).
\label{P1P2}
\end{eqnarray}
As before, $\omega_{\vec k}=2S\varepsilon_{\vec k}$.
The longitudinal correlations expanded in inverse powers
of $2S$ can be expressed as a sum of two contributions 
\beq
F^{\zeta \zeta}({\vec k}, \omega) = F^{\zeta \zeta}_{0}({\vec k},
\omega)+F^{\zeta \zeta}_{1}({\vec k}, \omega), 
\eeq
where
\begin{eqnarray}\label{2M}
F^{\zeta \zeta}_{0}({\vec k}, \omega)=-\frac{1}{2 N} \sum_{{\vec q}}
\left( u_{{\vec q}} v_{{\vec k} -{\vec q}} + v_{{\vec
q}} u_{{\vec k}-{\vec q}} \right)^2
% \nonumber \\ &\times& 
\left( \frac{1}{ \omega_{\vec q} + \omega_{{\vec k}-{\vec
q}} - \omega - i \eta}+ \frac{1}{ \omega_{\vec q} +
\omega_{{\vec k}-{\vec q}} + \omega - i \eta} \right),
\end{eqnarray}
\begin{eqnarray}\label{2M1}
F^{\zeta \zeta}_{1}({\vec k}, \omega)&=& \frac{1}{2 S} \left\{ \left(
P^{(1)}({\vec k}, \omega) \right)^2 
\left[ G_{11}({\vec k}, \omega)+ G_{22}({\vec k}, \omega)
+2 G_{12}({\vec k}, \omega) \right]
+ \left( P^{(2)}({\vec k},\omega ) \right)^2 \cdot \right. \nonumber
\\ & & \left.
\left[ G_{11}({\vec k}, \omega)+ G_{22}({\vec k}, \omega)
-2 G_{12}({\vec k}, \omega) \right] 
+2 P^{(1)}({\vec k}, \omega) P^{(2)}({\vec k}, \omega) 
\left[ G_{11}({\vec k}, \omega)- G_{22}({\vec k}, \omega)\right]
\right\}.
\end{eqnarray}
\end{widetext}
Note that the term $F^{\zeta \zeta}_{1}({\vec k}, \omega)$ is
formally of order $1/(2S)^2$ compared to the leading one magnon
contributions Eq. (\ref{1M}). However, it is kept because it
contains the self-energy corrections to Eq. (\ref{2M}) describing the
two magnon continuum. 

Using the set of formulas above, we have calculated the
scattering intensities as a function of frequency for 
various trajectories in $\vec k$ and $\omega$ space, corresponding
to the measured spectra.\cite{coldea} In general our results agree with 
those of Veillette et al.\cite{veillette} except for small differences
due to their use of self-consistent energy denominators in the self-energies 
and our use of shifted ${\vec Q}$ values in the anharmonic Green's functions.
Veillette et al. restricted their attention to values of $J$, $J^\prime$ and 
$D$ given by Eqs. (\ref{J})-(\ref{D}).  In the discussion that follows, we will
also consider the consequences of varying the $D$ parameter, which both shifts 
certain modes and also controls the size of quantum fluctuations, as well as
varying the overall energy scale, $J$.

We focus our attention on two
trajectories in the $(\vec k, \omega)$ space
whose parameterizations correspond to the scans {\rm G} and {\rm J} 
of Ref.~\onlinecite{coldea},
\begin{eqnarray}\label{scans}
&G: \quad k_x=0.5, & k_y=1.53-0.32\omega-0.1\omega^2 \nonumber\\
&J: \quad k_x=0.47,& k_y=1.0-0.45\omega.
\end{eqnarray}
In Eq. (\ref{scans}), $k_x$ and $k_y$ are written in 
the units of $2\pi$ and $2\pi/\sqrt{3}$ respectively, while
the energy is measured in meV.
Since the momentum transfer lies in the $x$-$z$ plane, 
$p_x=p_y=1$ for both of those scans. It is important also to take 
into account the finite energy and momentum resolution of the
scattered neutrons. The finite energy resolution 
$\Delta \omega \approx 0.016$ meV leads to a small Lorentzian 
broadening of the one-magnon peaks. However, the effect of smearing due 
to the finite momentum resolution $\Delta k/2\pi=0.085$ 
is much more pronounced and considerably broadens
the structure of the {\rm G} and {\rm J} scans.
What is special about the G and J scans is that they both exhibit 
sharp low-energy modes with energies around 0.1 meV.  As we shall see,
the energies of these peaks are quite sensitive to the exact value
of the DM interaction $D$.

\begin{figure}
\includegraphics[width=\columnwidth]{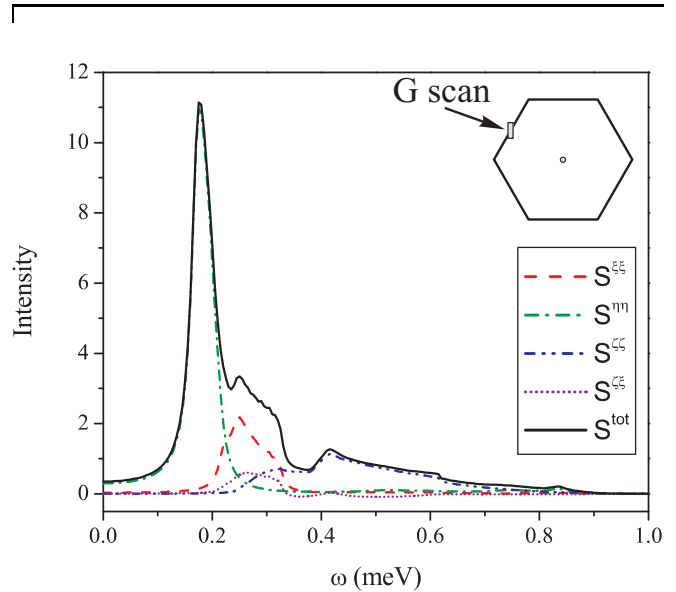}
\caption{(Color Online)
Intensity of the scattered neutrons for scan G in Eq. (\ref{scans}), in 
the presence of Dzyaloshinskii-Moria interaction $D=0.02$ meV. The 
energy and momentum resolutions are taken to be
$\Delta \omega =0.016$ meV and $\Delta k=0.085$ respectively. The
thick solid line is the total scattering intensity, Eq.~(\ref{totint}).
Other lines represent the various contributions appearing 
in Eqs.~(\ref{Syy})-(\ref{1M}).
}\label{stG}
\end{figure}

In Fig.~\ref{stG}, we plot the results of scattering intensity 
calculations for scan G, a data set which was analyzed in detail by 
Coldea et al.\cite{coldea} In particular, it was claimed that the
high-energy tail in this scan has too much intensity, by nearly
an order of magnitude, to be
explained by linear spin-wave theory.  
Our calculations were done using the bare couplings $J$ and
$J^{\prime}$, and the Dzyaloshinskii-Moriya interaction was set 
equal to $D=0.02$ meV, the value obtained from experiments in high
magnetic field.\cite{coldea2} The theory predicts three peaks, 
one corresponding to the principal mode $\varepsilon_{\vec k}$ and the 
other two, reflecting the presence of the shifted secondary spin waves 
at $\varepsilon_{\vec Q \pm \vec k}$. For the non-renormalized couplings 
$J$ and $J^{\prime}$, Eq. (\ref{spectrum}) for the bare
spectrum predicts that the principal peak should be at 
$0.22$ meV while the secondary peaks are very 
close to each other and located approximately at energy $0.28$ meV. 
Inclusion of self-energies to order $1/(2S)$ renormalizes the 
position of the main peak down to $0.18$ meV. The principal one-magnon 
peak should lie below a two-magnon scattering continuum starting
exactly at the location of the principal peak.
However, the presence of self-energies in the Green's functions, 
leads to a shift in positions of their poles, and to a 
finite gap between the principal peak and the start of the 
two magnon scattering continuum. This result is an artifact of 
truncating the perturbation expansion at a finite order, 
as discussed by Veillette et al.\cite{veillette} We
also observe that the two secondary peaks are
so close to each other that it is difficult to distinguish
them, once the finite frequency and momentum resolution is 
fully incorporated into the calculations. Experimentally, the main peak is 
observed at a somewhat smaller energy of $0.107$ meV, while the secondary 
ones are located around $0.25$ meV.\cite{coldea} 
Fig.~\ref{stG} also shows the sign and magnitude of the various 
contributions to the total intensity. It is seen that the two-magnon 
part, integrated over the full range of frequencies, carries a 
considerable overall weight compared to the one-magnon contribution. 
At the same time, the term mixing the transverse and longitudinal 
fluctuations can have both signs, and being small, does not influence 
the intensity profile very much. 

Why is the two-magnon intensity in our calculation so much larger, 
relative to the principal peak than in the analysis of Coldea et al.?
The reason is the factor $c^2_y$ which renormalizes the out-of-plane 
scattering and which is basically due to the reduction of the local 
moment by quantum fluctuations.  As noted above, theory predicts a much 
smaller local moment than the surprisingly large value obtained from 
experiment.\cite{coldea3}  According to our calculations, the moment 
reduction  increases with decreasing $D$, and the ordered moment vanishes 
for $D < 0.003$ meV.

\begin{figure}
\includegraphics[width=\columnwidth]{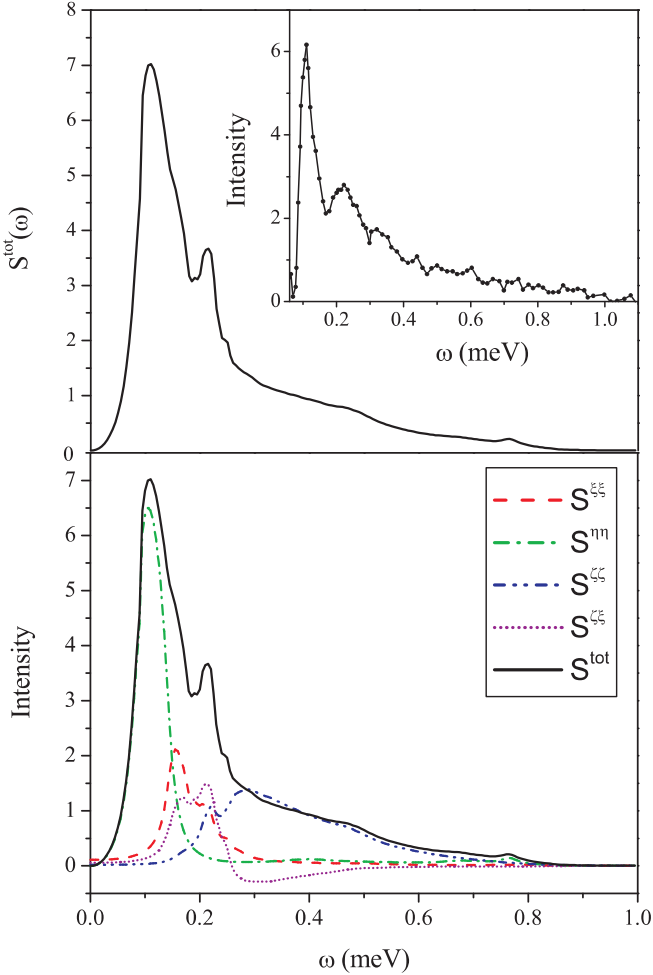}
\caption{(Color Online)
Intensity of the scattered neutrons corresponding to scan G, 
for the value of  Dzyaloshinskii-Moria interaction
$D=0.01$ meV. {\it Upper panel:} The total structure factor with the
energy (momentum) resolution $\Delta E=0.016$ meV ($\Delta
k=0.085$). {\it Inset:} Experimental data from
Ref. \onlinecite{coldea}. {\it Lower panel:} Different components which
contribute to the structure factor (see  Eqs.~(\ref{Syy})-(\ref{1M}) 
in the text). 
}
\label{stG05}
\end{figure}

In Fig.~\ref{stG05}, we plot the theoretical result for scan G, but 
this time with a reduced DM interaction $D=0.01$ meV. The comparison 
of our calculations with the experimental measurements reported in 
Ref.~\onlinecite{coldea}, is displayed in the upper panel. 
We see that for this value of $D$, the position of the principal 
magnon peak is shifted towards smaller energies. 
The gap between the main peak and the energy where the continuum
appears, is also decreased; there is further broadening of the principal 
peak and more intensity in the two-magnon continuum. It appears that the
value $D=0.01$ meV provides a better explanation of the form of
the total scattering intensity for scan G than $D=0.02$ meV. 
Comparison of the plots for a range of values of $D$, shows that 
the scattering intensity is very sensitive to the strength of the 
DM interaction.  [Note that the DM interaction was not included in the
analysis of Ref.~\onlinecite{coldea}.] Finally, we observe that 
even this smaller value of $D = 0.01$ meV does not account for all 
of the high energy scattering which seems to extend up to 1 meV, 
suggesting additional correlations at
high energy which are not contained in the theory.

\begin{figure}
\includegraphics[width=\columnwidth]{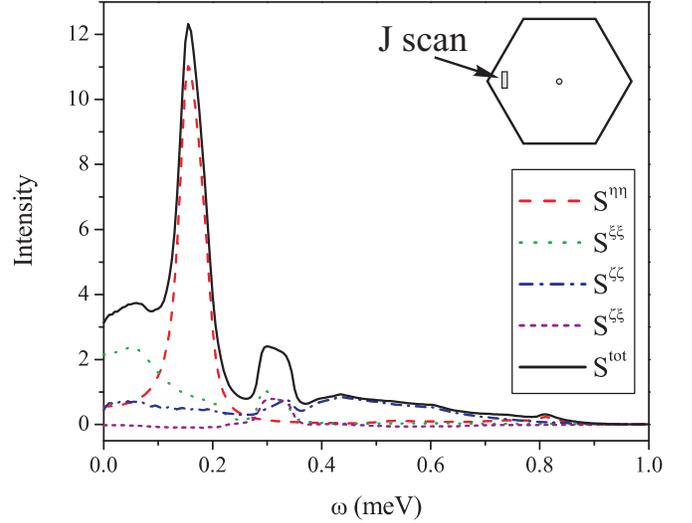}
\caption{(Color Online)
Intensity of the scattered neutrons corresponding to scan J,
Eq.~(\ref{scans}), in the presence of Dzyaloshinskii-Moria interaction
$D=0.02$ meV. The  different components are displayed as 
in Figs.~\ref{stG},~\ref{stG05}
}\label{stJ}
\end{figure}

Next, we examine scan J which probes the vicinity of the ordering 
wave vector.\cite{coldea}  Inelastic scattering at this point should show 
a primary peak at the wave vector ${\vec k_J} \approx {\vec Q}$ and two 
secondary modes, corresponding to momentum transfer ${\vec k_J} \pm {\vec Q}$. 
The primary mode, which is prominent in the theoretical spectrum, directly 
measures the strength of the DM interaction, since its energy is proportional
to $\sqrt{DJ}$. The experimental data for point J, 
shown only in the inset of 
Fig. 5G in the paper of Coldea and co-workers,\cite{coldea} do not extend to 
low enough energy to display this peak.  Nevertheless the position of the peak,
at an energy of 0.10(1) meV, is listed in Table 1 of that paper.  
This value of the mode at the J point is noticeably smaller than 
anharmonic spin-wave theory would predict for $D=0.02$ meV, suggesting that the
effective value of the DM interaction constant 
$D$ is smaller than 0.02 meV. Since the theory,  
for $D=0.02$ meV, 
gives a peak at 0.15 meV, a fit to the measured peak at 0.1 meV would 
suggest a 
value of $D$ in the range of 0.008 to 0.01 meV.  
Of the two secondary peaks at the J point, one has an energy close to zero, 
corresponding to the shifted zone-center mode, while the other lies at 
higher energy, just above 0.3 meV.

\section{Discussion and Conclusions}

The main motivation for the work presented in this paper was to 
see if a relatively 
conventional anharmonic spin-wave calculation could account for 
neutron scattering measurements on ${\rm Cs_2 Cu Cl_4}$.  Our
conclusion is a qualified ``yes''. 

Without any calculations, one knows that the observed excitations 
are from an ordered state exhibiting magnetic Bragg peaks, and hence that the 
lowest-lying excitations must be spin waves. However, both linear and anharmonic
spin-wave theory suggest a substantial moment reduction for a broad range
around the expected values of $D/J$ and $J^\prime/J$.  Further experiments 
would be useful to verify the amount by which the moment is reduced, since the 
present result\cite{coldea3} is inconsistent both with anharmonic spin-waves and with 
close proximity to a spin-liquid state.

At low energies we find that we can fit the sharp spin-wave
features of the data, albeit with a somewhat smaller value of the DM 
interaction constant, $D$, than was obtained from high field measurements of a 
fully polarized state.
We also find from theoretical calculations that, without a
sufficiently 
large DM interaction,
long range order is destroyed by quantum fluctuations.  However, 
the value of $D$ that best fits theory to experiment is well within 
the range in which order is stable, at least for 
$T=0$. 
 Additional experiments, including measurements of the renormalized $D$
in zero field, might
clarify this point. (We note in passing that
a value of $J$ about 30-40\% larger than given by Eq. (\ref{J}), for fixed
$J^\prime/J$, would provide a much improved fit to the higher energy 
one-spin-wave features.)

Our theory yields a considerable amount of continuous  
two-spin-wave scattering.  However the 
amount of this scattering, although
readily measurable, is somewhat less than is observed experimentally.  
Furthermore, calculation of the relative 
size of one- and two-spin-wave features is complicated by their 
sensitivity to the size of $D$.  
That is, a smaller $D$ suppresses one-spin-wave scattering and makes 
two-spin-wave continuous scattering relatively more conspicuous.  
Quantum fluctuations also broaden the one-spin-wave features via 
lifetime effects.  However, it is difficult 
to assess, from the data alone, how much of the observed broadening of the 
principal peaks is due to lifetimes and how much is due to 
low momentum resolution as was emphasized by Veillette et al.\cite{veillette} 
The theory suggests though that experiments with higher momentum resolution
are likely to reveal considerably sharper low energy spin-wave peaks.

With regard to the shape of the broad scattering at higher energies, 
we are inclined toward a picture of 
scattering by quasi-one-dimensional spinons.  With a bare interchain 
coupling of 0.128 meV, which we know to be considerably renormalized 
downward by quantum fluctuations, one might 
expect that, at energies above about
0.3 meV, the system would behave like a collection of
uncoupled chains. In this sense we would expect
the high-energy response of the system to be better described by
one-dimensional spinons than by a low-order spin-wave perturbation
calculation such as the one that we have studied.

\begin{acknowledgments}
This project has been supported by the Natural Sciences and Engineering Research
Council (NSERC) of Canada and by the Canadian Institute for Advanced Research (CIAR).  The authors
gratefully acknowledge valuable discussions and correspondence with Martin Veillette,
who also provided access to a particularly useful computer program. Some of the research for this 
paper was done at the Aspen Center for Physics.
\end{acknowledgments}

\appendix
\section{$1/(2S)^2$ Correction to the Ground State Energy and 
Ordering Wave Vector }

The ground state energy correction at the order $1/(2S)^2$, 
denoted here as $E_G^{(2)}(Q)$, consists of two parts,
\beq
E_G^{(2)}(Q)= {\cal E}_3 (Q) +{\cal E}_4 (Q).
\eeq
${\cal E}_4(Q)$ is obtained simply by substituting the 
Bogoliubov transformation Eq. (\ref{bogol}) into ${\cal H}^{(4)}$ and 
performing the normal ordering in the terms containing two pairs of 
quasiparticle operators $c$ and $c^{\dagger}$. In this Appendix,
we present the results of calculations in which the 
Dzyaloshinskii-Moria interaction $D=0$, so that
\begin{eqnarray}
\label{e4}
{\cal E}_4 (Q) &= \displaystyle\frac{1}{2N}  \sum_{\vec k} 
\left[ \varepsilon_{\vec k} \left( 1-
\frac1N \sum_{\vec p} \frac{ A_{\vec p}}{\varepsilon_{\vec p}} 
\right)
-\displaystyle\frac{A_{\vec k}}{2} \right]+\nonumber\\
& \displaystyle\frac{1}{N^2}\sum_{\vec k, \vec p} 
( A_{\vec k -\vec p} -  B_{\vec k -\vec p} )
\displaystyle\frac{A_{\vec k}A_{\vec p} + B_{\vec k}B_{\vec p}}
{4\varepsilon_{\vec k} \varepsilon_{\vec p}}.
\end{eqnarray}

${\cal E}_3(Q)$ ensues due to the key role of the
cubic interactions (\ref{H3}) in the total Hamiltonian. It is 
technically advantageous first to express (\ref{H3}) in terms of the 
quasiparticle $c$-operators, so that
\begin{eqnarray}\label{H3new}
{\cal H}^{(3)} = \displaystyle\frac{i}{2} &\displaystyle\sqrt{\frac{S}{2N}}  
\displaystyle\sum_{{\vec 1},{\vec 2},{\vec 3}} 
 \left\{ \displaystyle\frac13 \Gamma_1 ({\vec 1},{\vec 2},{\vec 3})
\left( 
c^{}_{\vec 3} c^{}_{\vec 2} c^{}_{\vec 1} - c^{\dagger}_{\vec 1}  
c^{\dagger}_{\vec 2} c^{\dagger}_{\vec 3}
\right) 
\right.\nonumber\\
&\left. + \Gamma_2 ({\vec 1},{\vec 2};{\vec 3}) \left(
c^{\dagger}_{\vec 3} c^{}_{\vec 2} c^{}_{\vec 1}-
c^{\dagger}_{\vec 1}  c^{\dagger}_{\vec 2} c^{}_{\vec 3} 
\right) \right\},
\end{eqnarray}
\begin{eqnarray}\label{Gamma1}
\Gamma_1 ({\vec 1},{\vec 2},{\vec 3})& = &\left( {C_{\vec 1}} 
+{C_{\vec 2}} \right) \left( 
u_{\vec 1} u_{\vec 2} v_{\vec 3} + v_{\vec 1} v_{\vec 2} u_{\vec 3} 
\right)\nonumber\\
&+ & \left( {C_{\vec 1}} +{C_{\vec 3}} \right) \left( 
u_{\vec 1} u_{\vec 3} v_{\vec 2} + v_{\vec 1} 
v_{\vec 3} u_{\vec 2} \right) \nonumber\\
&+ & \left( {C_{\vec 2}} +{C_{\vec 3}} \right) \left( 
u_{\vec 2} u_{\vec 3} v_{\vec 1} +
v_{\vec 2} v_{\vec 3} u_{\vec 1} \right),
\end{eqnarray}
\begin{eqnarray}\label{Gamma2}
\Gamma_2 ({\vec 1},{\vec 2};{\vec 3})& = &( {C_{\vec 1}} +
{C_{\vec 2}} ) (u_{\vec 1} u_{\vec 2} u_{\vec 3} + 
v_{\vec 1} v_{\vec 2} v_{\vec 3} )\nonumber\\
&+ & ( {C_{\vec 1}} -{C_{\vec 3}} ) (v_{\vec 1} 
u_{\vec 2} u_{\vec 3} + u_{\vec 1} v_{\vec 2} v_{\vec 3} )
\nonumber\\
&+ & ( {C_{\vec 2}} -{C_{\vec 3}} ) (u_{\vec 1} 
v_{\vec 2} u_{\vec 3} + v_{\vec 1} u_{\vec 2} v_{\vec 3} ).
\end{eqnarray}
At $T=0$, the only contribution
to ${\cal E}_3 (Q)$ is given by the 'sunrise' diagram containing two vertices
$\Gamma_1(\vec q, \vec p, -\vec q -\vec p)$. Using the formulas
(\ref{uv}), we arrive at the following expression for ${\cal E}_3$:
\begin{widetext}
\begin{eqnarray}\label{e3}
{\cal E}_3 (Q) & = -\displaystyle\frac{1}{24N^2} \sum_{\vec p, \vec q}
\displaystyle\frac{\left[ \Gamma_1 (\vec q, \vec p, -\vec q -\vec p)
\right]^2}{\varepsilon_{\vec p}+\varepsilon_{\vec q}+
\varepsilon_{\vec p+\vec q}}
= -\displaystyle\frac{1}{16N^2}\sum_{\vec p, \vec q}
\displaystyle\frac{1}{(\varepsilon_{\vec p}+\varepsilon_{\vec q}+
\varepsilon_{\vec p+\vec q})} \left\{  C_{\vec p +\vec q}^2
\left( \frac{A_{\vec p +\vec q} +B_{\vec p +\vec q}}
{\varepsilon_{\vec p+\vec q}} \right) 
\left( \frac{A_{\vec p} A_{\vec q}+ B_{\vec p} B_{\vec q}}
{\varepsilon_{\vec p}\varepsilon_{\vec q}} -1 \right)
\right. \nonumber\\
&\left. +  {C_{\vec p}}{C_{\vec q}} \left[
\displaystyle\frac{ ( A_{\vec p +\vec q} +B_{\vec p +\vec q} ) 
( A_{\vec p} +B_{\vec p} ) ( A_{\vec q} +B_{\vec q} )}
{\varepsilon_{\vec p+\vec q}\varepsilon_{\vec q}\varepsilon_{\vec p}}
+ \displaystyle\frac{A_{\vec p +\vec q} -B_{\vec p +\vec q}}
{\varepsilon_{\vec p+\vec q}}
-\displaystyle\frac{A_{\vec p} +B_{\vec p}}{\varepsilon_{\vec p}}
-\displaystyle\frac{A_{\vec q} +B_{\vec q}}{\varepsilon_{\vec q}}
\right] \right\}.
\end{eqnarray}
\end{widetext}

Numerical calculations show that for the interaction constants given
by Eqs. (\ref{J}),
\beq 
{\cal E}_3 (Q_0)/J=-0.0425,
\eeq
\beq
{\cal E}_4 (Q_0)/J=\quad 0.0103,
\eeq
which are quite small. The $1/(2S)^2$ correction to $Q_0$ can be 
computed by substituting the expansion
\beq
Q=Q_0 +\displaystyle\frac{\Delta Q^{(1)}}{2S}+
\displaystyle\frac{\Delta Q^{(2)}}{(2S)^2}
\eeq
into the series for the ground state energy
\beq
E_G(Q)= S^2 J_{\vec Q} + S E_G^{(1)}(Q)+ E_G^{(2)}(Q),
\eeq
and expanding its minimum in powers of $1/(2S)$. As a result,
\begin{eqnarray}\label{Q2}
\Delta Q^{(2)} & =  -\left[ 
\displaystyle\frac{\partial^2 J_{\vec Q}}{\partial Q^2}
\right] ^{-1} \cdot\left[ \displaystyle\frac12 
\displaystyle\frac{\partial^3 J_{\vec Q}}{\partial Q^3} \cdot 
(\Delta Q^{(1)})^2 
\right.\nonumber\\
&\left.+2 \displaystyle\frac{\partial^2 E_G^{(1)}(Q)}{\partial Q^2} \cdot
(\Delta Q^{(1)})
+4 \displaystyle\frac{\partial E_G^{(2)}(Q)}{\partial Q}
\right]
\Biggr\vert \sb{Q_0}.
\end{eqnarray}
The calculations of derivatives over $Q$ in expressions entering 
Eqs. (\ref{EG1}), (\ref{e3}), (\ref{e4}) are lengthy but
straightforward. Omitting the corresponding cumbersome final expressions, we
state here that as a result of numerical integration:
\beq
\Delta Q^{(2)}/2\pi \approx -0.011.
\eeq
The value of the ground state energy at its minimum ($S=1/2$) 
\begin{widetext}
\beq\label{EGmin}
E_{min}/J = {1\over J} \left[ S^2 J_{\vec Q} + S E_G^{(1)}(Q)+ E_G^{(2)}(Q)
-\displaystyle\frac18 \displaystyle\frac{\partial^2 J_{\vec Q}}
{\partial Q^2} \cdot (\Delta Q^{(1)})^2 \right] 
\Biggr\vert \sb{Q_0} =  -0.459.
\eeq
Note that the last term in Eq. (\ref{EGmin}) is always negative and 
arises solely due to the shift in the ordering wave vector $\Delta
Q^{(1)}$. Numerically however, this term is tiny, even compared to the
second order correction $E_G^{(2)}(Q_0)$ that itself gives a small 
contrbution to the total energy Eq. (\ref{EGmin}).
\end{widetext}


\begin{thebibliography}{99}
\frenchspacing
\bibitem{senthil1} See T. Senthil, cond-mat/0411275, and references therein.
\bibitem{pwa} P. W. Anderson, Mater. Res. Bull. {\bf 8}, 153 (1973); 
P. Fazekas and P. W. Anderson, Philos. Mag. {\bf 30}, 23 (1974).
\bibitem{kivelson} S. A. Kivelson, D. S. Rokhsar, and J. P. Sethna,
 Phys. Rev. B {\bf 35}, 8865 (1987).
\bibitem{luttinger} A. O. Gogolin, A. A. Nersesyan and A. M. Tsvelik,
{\it Bosonization in Strongly Correlated Systems} (Cambridge
University Press, Cambridge, 1999).
\bibitem{FQHE} See, e. g., {\it Perspectives in Quantum Hall Effects:
Novel Quantum Liquids in Low-Dimensional Semiconductor Structures },
Ed. by  S. Das Sarma and A. Pinzcuk (Wiley, New York, 1996).
\bibitem{spinliq} R. Moessner and S.L. Sondhi, Phys. Rev. Lett. 
{\bf 86}, 1881 (2001).
\bibitem{spinliq1} C. Nayak and K. Shtengel, Phys. Rev. B {\bf 64}, 
064422 (2001).
\bibitem{balents1} L. Balents, M. P. A. Fisher, and C. Nayak,
Int. J. Mod. Phys. B {\bf 12}, 1033 (1998).
\bibitem{balents2} L. Balents, M. P. A. Fisher, and C. Nayak,
Phys. Rev. B {\bf 60}, 1654 (1999).
\bibitem{balents3} L. Balents, M. P. A. Fisher, and C. Nayak,
Phys. Rev. B {\bf 61} 6307 (2000).
\bibitem{demler}
E. Demler, C. Nayak, H.-Y. Kee, Y. B. Kim, and T. Senthil, Phys. Rev. 
B {\bf 65}, 155103 (2002).
\bibitem{senthil}
T. Senthil and O. Motrunich, Phys. Rev. B {\bf 66}, 205104 (2002);
O. I. Motrunich and T. Senthil,Phys. Rev. Lett. {\bf 89}, 277004 (2002).
\bibitem{spinliq2} L. Balents, M.P.A. Fisher and S.M. Girvin, 
Phys. Rev. B {\bf 65}, 224412 (2002).
\bibitem{coldea} R. Coldea, D. A. Tennant and Z. Tylczynski,
Phys. Rev. B {\bf 68}, 134424 (2003).
\bibitem{coldea1} R. Coldea, D. A. Tennant and Z. Tylczynski, 
Phys. Rev. Lett. {\bf 86}, 1335 (2001).
\bibitem{series} Z. Weihong, R. H. McKenzie and R. R. P. Singh, 
Phys. Rev. B {\bf 59}, 14367 (1999).
\bibitem{series2} W. Zheng, J. O. Fjaerestad, R. R. P. Singh, R. H. McKenzie, and R. Coldea, cond-mat/0506400.
\bibitem{largen} C. H. Chung, J. B. Marston and R. H. McKenzie, J. Phys.: 
Condens. Matter {\bf 13}, 5159 (2001); S. Takei, C. H. Chung
and Y. B. Kim, Phys. Rev. {\bf B 70}, 104402 (2004). 
\bibitem{slaveboson} Y. Zhou and X. G. Wen, cond-mat/0210662.
\bibitem{slaveboson1} C.-H. Chung, K. Voelker, and Y. B. Kim,
Phys. Rev. B {\bf 68}, 094412 (2003).
\bibitem{merino} J. Merino, Ross H. McKenzie, J. B. Marston and
C. H. Chung, J. Phys.: Condens. Matter {\bf 11} , 2965 (1999).
\bibitem{alicea}J. Alicea, O. I. Motrunich, M. P. A. Fisher, 
cond-mat/0512427.
\bibitem{veillette}M. Y. Veillette, A. J. A. James, and F. H. L. Essler,
Phys. Rev. B {\bf 72}, 134429, (2005).
\bibitem{coldea2} R. Coldea, D. A. Tennant, K. Habicht, P. Smeibidl,
C. Wolters and Z. Tylczynski,  Phys. Rev. Lett. {\bf 88}, 
137203(2002).
\bibitem{holstein} T. Holstein and H. Primakoff, Phys. Rev. {\bf 58}, 1098
(1940).
\bibitem{tokiwacold} Y. Tokiwa, T. Radu, R. Coldea, H. Wilhelm,
Z. Tylczynski, and F. Steglich, cond-mat/0601272 (2006).
%\bibitem{chubukov}
%See e.g. A. V. Chubukov. Sov. Phys. JETP {\bf 62}(4), 763 (1985).
\bibitem{lovesey} 
S. W. Lovesey, in {\it Theory of Neutron Scattering from Condensed Matter},
(Clarendon, Oxford, England, 1987).
\bibitem{abrikos} See, e.g., A. A. Abrikosov, L. P. Gorkov \& I. E.
Dzyaloshinski. {\it Methods of Quantum Field Theory in Statistical Physics}, 
(Dover, New York, 1975)
\bibitem{ohyama} T. Ohyama and H. Shiba, J. Phys. Soc. Jpn. {\bf 62},
3277 (1993).
\bibitem{chubuk} A. V. Chubukov, S. Sachdev, and T. Senthil, 
J. Phys.: Condens. Matter {\bf 6}, 8891 (1994).
\bibitem{coldea3} R. Coldea, D.A. Tennant, R.A. Cowley, D.F. McMorrow,
B. Dorner, and Z. Tylczynski, J. Phys.: Condens. Matter {\bf 8}, 7473 (1996).
\end{thebibliography}
\end{document}